\documentclass{article}
\usepackage{spconf,amsmath,amssymb,graphicx,url}
\usepackage{tikz}
\usepackage{comment}
\usepackage[
    backend=biber,
    style=ieee,
    maxbibnames=3,
    maxcitenames=3,
    doi=false,isbn=false,url=false,eprint=false
]{biblatex} 

\newcommand{\Fig}[1]{Figure~\ref{fig:#1}} % refer to figure
\newcommand{\Table}[1]{Table~\ref{tab:#1}} % refer to table
\newcommand{\drawfig}[4]{ % draw figure 
    \begin{figure}[#1]
    \centering 
    \vspace{0mm}
    \includegraphics[width=#2,clip]{#3.pdf} 
    \vspace{-3.5mm}
    \caption{#4}
    \label{fig:#3}
    \vspace{-4mm}
    \end{figure}
}

\DeclareSourcemap{
	\maps[datatype=bibtex, overwrite=true]{
		\map{
			\step[fieldsource=booktitle,
			match=\regexp{.*Interspeech.*},
			replace={Proc. Interspeech}]
			\step[fieldsource=journal,
			match=\regexp{.*INTERSPEECH.*},
			replace={Proc. Interspeech}]
			\step[fieldsource=booktitle,
			match=\regexp{.*ICASSP.*},
			replace={ICASSP}]
			\step[fieldsource=booktitle,
			match=\regexp{.*icassp_inpress.*},
			replace={ICASSP (in press)}]
			\step[fieldsource=booktitle,
			match=\regexp{.*International.*Conference.*on.*Acoustics,.*Speech.*and.*Signal.*Processing.*},
			replace={ICASSP}]
			\step[fieldsource=booktitle,
			match=\regexp{.*International.*Conference.*on.*Learning.*Representations.*},
			replace={ICLR}]
			\step[fieldsource=booktitle,
			match=\regexp{.*International.*Conference.*on.*Machine.*Learning.*},
			replace={ICML}]
			\step[fieldsource=booktitle,
			match=\regexp{.*Automatic.*Speech.*Recognition.*and.*Understanding.*},
			replace={Proc. ASRU}]
			\step[fieldsource=booktitle,
			match=\regexp{.*Spoken.*Language.*Technology.*},
			replace={Proc. SLT}]
			\step[fieldsource=booktitle,
			match=\regexp{.*Speech.*Synthesis.*Workshop.*},
			replace={Proc. SSW}]
			\step[fieldsource=booktitle,
			match=\regexp{.*workshop.*on.*speech.*synthesis.*},
			replace={Proc. SSW}]
			\step[fieldsource=booktitle,
			match=\regexp{.*Advances.*in.*neural.*information.*processing.*},
			replace={Proc. NIPS}]
			\step[fieldsource=booktitle,
			match=\regexp{.*Advances.*in.*Neural.*Information.*Processing.*},
			replace={Proc. NIPS}]
			\step[fieldsource=booktitle,
			match=\regexp{.*Workshop.*on.* Applications.* of.* Signal.*Processing.*to.*Audio.*and.*Acoustics.*},
			replace={Proc. WASPAA}]
			\step[fieldsource=booktitle,
			match=\regexp{.*International.*Conference.*on.*Language.*Resources.*and.*Evaluation.*},
			replace={Proc. LREC}]
			\step[fieldsource=journal,
			match=\regexp{.*Spontaneous.*Speech.*Processing.*and.*Recognition},
			replace={Proc. SSPR}]
			\step[fieldsource=publisher,
			match=\regexp{.+},
			replace={{}}]
			\step[fieldsource=month,
			match=\regexp{.+},
			replace={{}}]
			\step[fieldsource=location,
			match=\regexp{.+},
			replace={{}}]
			\step[fieldsource=address,
			match=\regexp{.+},
			replace={{}}]
			\step[fieldsource=organization,
			match=\regexp{.+},
			replace={{}}]
			\step[fieldsource=doi,
			match=\regexp{.+},
			replace={{}}]
			\step[fieldsource=url,
			match=\regexp{.+},
			replace={{}}]
			\step[fieldsource=editor,
			match=\regexp{.+},
			replace={{}}]
		}
	}
} 

\title{JTubeSpeech: corpus of Japanese speech collected from YouTube\\for speech recognition and speaker verification}
\name{Shinnosuke Takamichi$^1$, Ludwig Kürzinger$^2$, Takaaki Saeki$^1$, Sayaka Shiota$^3$, Shinji Watanabe$^4$\thanks{We would like to thank Hiromasa Fujihara and the GigaSpeech team, especially Guoguo Chen and Shuzhou Chai for their valuable comments. This work used the HLTCOE cluster at Johns Hopkins University and the Extreme Science and Engineering Discovery Environment (XSEDE) \cite{towns2014xsede}, which is supported by National Science Foundation grant number ACI-1548562. Specifically, it used the Bridges system \cite{nystrom2015bridges}, which is supported by NSF award number ACI-1445606, at the Pittsburgh Supercomputing Center (PSC). This work is also supported by JSPS KAKENHI 19K20271, 21H04900, 21H05054, JST Moonshot R\&D Grant Number JPMJPS2011, ROISDS-JOINT (030RP2021) to S. Shiota and the SECOM Science and Technology Foundation.}}
\address{$^1$The University of Tokyo, Japan, $^2$Technical University of Munich, Germany,\\$^3$Tokyo Metropolitan University, Japan, $^4$Carnegie Mellon University, USA}

\begin{document}
\ninept
\maketitle
\setlength{\tabcolsep}{1mm} 

% abstract
\begin{abstract} \vspace{-2.5mm}
    In this paper, we construct a new Japanese speech corpus called ``JTubeSpeech.''
    Although recent end-to-end learning requires large-size speech corpora, open-sourced such corpora for languages other than English have not yet been established.
    In this paper, we describe the construction of a corpus from YouTube videos and subtitles for speech recognition and speaker verification.
    Our method can automatically filter the videos and subtitles with almost no language-dependent processes.
    We consistently employ Connectionist Temporal Classification (CTC)-based techniques for automatic speech recognition (ASR) and a speaker variation-based method for automatic speaker verification (ASV).
    We build 1) a large-scale Japanese ASR benchmark with more than 1,300 hours of data and 2) 900 hours of data for Japanese ASV.
    % ASRU 
    %In this paper, we construct a new Japanese speech corpus called ``JTubeSpeech.''
    %With the recent development of end-to-end and self-supervised learning, developing a large speech corpus contributes to improving the performance of many speech-related tasks.
    %However, open-sourced modern-size speech corpora for languages other than English have not yet been established.
    %In this paper, we demonstrate the construction of a corpus from YouTube videos and subtitles for speech recognition and speaker verification.
    %From many videos and subtitles for developing the corpus, our method can automatically extract and refine the data with almost no language-dependent processes.
    %For example, for the automatic speech recognition (ASR) benchmark design, we consistently employ Connectionist Temporal Classification (CTC)-based techniques to refine subtitle timing and prune wrong utterances with minimal language-dependent processing.
    %Furthermore, we propose a speaker variation-based method for filtering utterances for automatic speaker verification (ASV).
    %Experimental evaluation shows that 1) 10,000 hours of data are collected from videos with manual subtitles, 2) we build a large-scale Japanese ASR benchmark with more than 1,300 hours of data, and 3) we also build 900 hours of data for Japanese ASV.
\end{abstract} \vspace{-1mm}

% keyword
\begin{keywords} 
    automatic speech recognition, automatic speaker verification, speech corpus, YouTube
\end{keywords}

% Section 1: Introduction
\vspace{-2mm}
\section{Introduction} \label{sec:introduction}
\vspace{-2mm}
Powered by the development of deep learning, significant progress has been made on various speech recognition tasks, e.g., automatic speech recognition (ASR)~\cite{george11hybridasr,graves2014e2easr,gulati20conformer} and automatic speaker verification (ASV)~\cite{variani14dvecs,snyder18xvector}. Due to data hungriness of deep learning, massive-size speech corpora have been constructed and published. It is desirable to build and publish speech corpora of all languages for decentralizing the speech technologies. However, corpora in languages other than English are very poor. For example, while several thousands of hours of corpora have been published in English and Chinese~\cite{cieri2004fisher,panayotov2015librispeech,ardila20commonvoice,oneill21spgispeech,liu06hkust,du2018aishell,fan19cnceleb}, similar-size corpora are very limited for other languages.

Japanese, the target language in this paper, is also this example. The CSJ corpus~\cite{maekawa00csj} is the most frequently used corpus for Japanese ASR, but its size is relatively small ($600$~ hours) compared to modern English corpora such as Common Voice~\cite{ardila20commonvoice} ($1,100$~hours). Also, the modern size of ASV corpora is more than $1,000$ hours~\cite{nagrani19voxceleb,fan19cnceleb}, but there is no open-sourced corpus for Japanese ASV.

To construct a large-scale corpus, several studies have collected text-audio pairs from videos~\cite{abuelhaija16youtube8m,nagrani19voxceleb,oneill21spgispeech,galvez2021peoplespeech}.
For example, YouTube provides videos in diverse genres, recording environments, speakers, and language accents. There is no doubt that such \textit{in-the-wild} data is useful for wide purpose of modern speech technology. Chen~\cite{chen2021gigaspeech} proposed a strategy to collect English videos for ASR use. Also, Fan~\cite{fan19cnceleb} manually selected Chinese celebrities and extracted videos for ASV use. Unlike these methods that require language-dependent and manual processes, this paper aims to develop a corpus with almost no language-dependent and no manual processes. The establishment of this method will be useful for building corpora of many languages, not limited to English and Chinese.

%%% ASRU %%%
% To construct a large-scale speech corpus, several studies have collected text-audio pairs from videos~\cite{abuelhaija16youtube8m,nagrani19voxceleb,oneill21spgispeech,galvez2021peoplespeech}. For example, YouTube provides videos in diverse genres, recording environments, speakers, and language accents. There is no doubt that such \textit{in-the-wild} data is useful for wide purpose of modern speech technology, e.g., ASR and ASV. Chen~\cite{chen2021gigaspeech} proposed a strategy to collect category-specific English videos for ASR use. They manually defined 24 categories and selected videos with manual (i.e., human-generated) captions. Also, Fan~\cite{fan19cnceleb} manually selected 1,000 Chinese celebrities and extracted videos of those people. Unlike these methods that require language-dependent and manual processes, this paper aims to develop a corpus with almost no language-dependent and no manual processes. The establishment of this method will be useful for building corpora of many languages, not limited to English and Chinese.

In this paper, we propose a mostly language-independent strategy to construct a speech corpus for both ASR and ASV, and then construct a Japanese speech corpus ``JTubeSpeech'' by using the proposed method. We first crawl YouTube in order to generate candidate audio-text pair data. For ASR, the subtitles are aligned to the audio using a CTC-based ASR model using CTC segmentation~\cite{ctcsegmentation}. This method calculates a confidence score to filter the audio-text pair~\cite{ctcsegmentation,zhang2021nemo,bakhturina2021hifi}. Unlike the conventional hidden Markov model (HMM) based cleaning \cite{trmal2017kaldi,guo2021recent}, this method does not require any language-dependent pre-processing thanks to the end-to-end framework. 
Also, for ASV, the audio can be filtered by calculating variation of speaker representation within a video. This paper applies the above techniques to a Japanese corpus as a case study. Experimental evaluation demonstrates 1) we designed a new Japanese ASR benchmark with more than 1,300 hours of training data and the official test sets, and 2) we also constructed 900 hours of Japanese ASV corpus. The contributions of this work are as follows:
    \vspace{-2mm}
    \begin{itemize} \leftskip -5mm \itemsep -1mm
        \item We construct a new modern-size corpus for Japanese ASR and ASV from YouTube videos. The video list is open-sourced in our project page\footnote{\url{https://github.com/sarulab-speech/jtubespeech}}.
        \item Our process is applicable to many languages with high reproducibility. The repository also contains the script, which has been extended to support multiple languages for the data collection.
    \end{itemize}
    \vspace{-2mm}

\vspace{-3mm}
\section{Corpus construction} \label{sec:data_collection}
\vspace{-2mm}

    \drawfig{t}{0.90\linewidth}{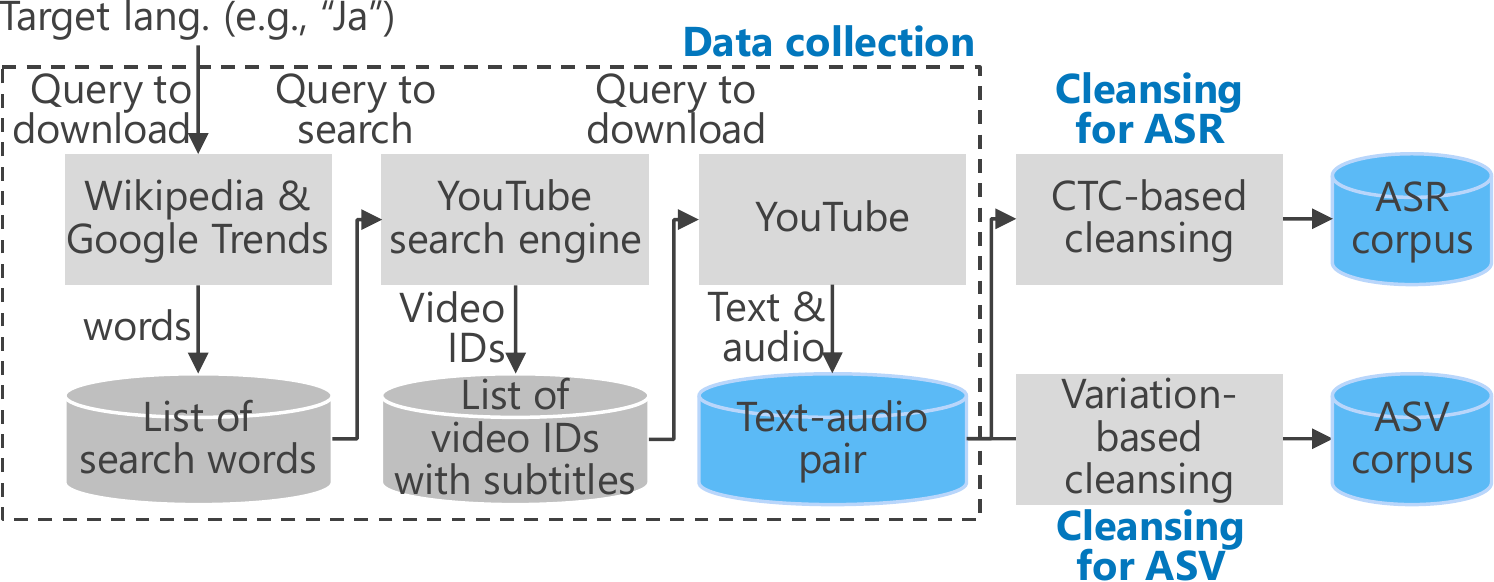}
    {Procedure of corpus construction.}
    All of our steps are performed with minimal language-dependent processing. \Fig{fig/overview_shortened} shows the procedure.
    
    %%% ASRU %%%
    % We describe steps to build our corpus. All the steps are performed with minimal language-dependent processing. \Fig{fig/overview_shortened} shows the procedure.

    \vspace{-2mm}
    \subsection{Data collection} \vspace{-1mm}
        % We obtain candidate data for the corpus, i.e., subtitles and audio data in the target language (Japanese in this paper) from YouTube. 
    
        \textbf{Creating search terms.} 
        The first step is to create search terms to be entered into the video search engine. We extract the target language's words with hyperlinks from HTML files of Wikipedia articles. Unlike Gigaspeech~\cite{chen2021gigaspeech}, categories are not specified, i.e., all articles are used. Also, we extract ``sudden-rise search terms'' in the past few years from Google Trends.%\footnote{\url{https://trends.google.com/}}. 
        
        \textbf{Obtaining video IDs that have subtitles.}
        Next, we retrieve video IDs that have subtitles. Entering a search word into the video search engine, we obtain a list of video ID candidates. Then, for each video, we retrieve whether the video has subtitles. In this paper, we use only manual subtitles but also make the list for automatic (i.e., machine-generated) subtitles. 
        
        \textbf{Downloading audio and caption.}
        Finally, we download the audio and manual subtitles from videos. Since the number of channels and sampling frequency of the audio file varies from video to video, we reformat the audio to 16~kHz-sampled monaural WAV format.
    
    \vspace{-2mm}
    \subsection{Specific cleansing for speech recognition} \label{sec:asr} \vspace{-1mm}
    
        The obtained audio data were already annotated with subtitles including timings.
        This dataset still included many bad samples, e.g., unspoken subtitles, English audio with Japanese subtitles, and other variations of audio--text mismatches.
        To sort-out these bad samples, we calculated a score of how well the audio segment fits to the subtitle and then filtered the utterances based on their score.
        Furthermore, as many subtitle timings were inaccurate, we fully re-aligned the subtitles to the audio.
        %
        %\paragraph{Scoring and CTC segmentation} \vspace{-1mm}
        %\label{sec:ctc_score_seg}
        To calculate a score and to re-align the subtitles to the audio, we use CTC segmentation~\cite{ctcsegmentation} as an alignment tool~\cite{ctcsegpython}. %\footnote{\url{https://github.com/lumaku/ctc-segmentation}}.
        CTC segmentation utilizes CTC log-posteriors to determine utterance timings in the audio given a ground-truth text.
        
        \textbf{Text pre--processing.}
        We apply minimal text pre--processing so that the ground truth text obtained from subtitles is composed of characters or tokens in the model dictionary.
        Numbers are replaced with their spoken equivalent using the \emph{num2words} Python library~\cite{num2words}. %\footnote{\url{https://github.com/savoirfairelinux/num2words}}.
        UTF-16 characters are mapped to the Japanese character set.
        Automated subtitles are detected and filtered out based on the average relative Levenshtein distance between subtitles.
        
        % Short description of CTC segmentation:
        \textbf{Alignment.}
        The onset and offset timings of an utterance are then estimated in three steps:
        (1) In a forward pass, transition probabilities are mapped into a trellis diagram of the ground-truth token sequence over all time steps.
        (2) Backtracking starts from the most probable timing of the last character and then determines the most probable path through the trellis.
        (3) A confidence score is derived for each utterance from the per-token probabilities in the trellis.
        The score is determined by the $L$ consecutive CTC output frames of the utterance with the lowest token probabilities; we chose $L=30$ that relates to approximately $1$s of audio.
        For a more in-depth description, readers could refer to \cite{ctcsegmentation}.
        We apply one further modification that helps with the alignment of fragmented subtitle text:
        In the original publication, the algorithm was configured to skip preambles by setting transition cost to zero for the token that marks the start of the first utterance.
        We extended this in our setup to all utterances in order to skip any unrelated audio segments.

        % explain CTC scoring here as a separate step - as it is referred to in Sec. 3.2.1
        \textbf{CTC scoring.}
        We also calculate a CTC score for each audio-text pair from its YouTube timings instead of fully aligning the subtitles to audio.
        For this, the audio segment to each subtitle is cut out to then derive its confidence score as described above.

        % confidence score length: Transformer: 768 --> 1.44s; RNN --> 0.95s
        \textbf{Cleaning.}
        Bad samples are then eliminated based on this confidence score.
        The confidence score provides an estimated log-space probability of how well the subtitles fit the audio data;
        this value is mainly influenced by the quality of ASR models, input data, and data pre-processing.
        Note that a score threshold $\theta$ of $-0.3$, used in one of our experiments, can be interpreted as a production probability of at least $75\%$ each second.
        
        \textbf{Inference of long audio files.}
        Alignment requires inference of the audio using the encoder and the CTC layer of a pre-trained ASR model.
        We found that many audio files are longer than three hours and their inference reaches a practical limitation:
        The memory complexity using a Transformer-based model is quadratic with audio length -- a device with 64 GB memory is at most able to infer $500$s of audio at once.
        Due to their rather linear memory complexity along longer audio data, RNN-based models require less memory and thus, it is possible to decode longer files.
        This can be scaled up to $2.7$h of audio until the encoding reaches a memory limit of the Pytorch toolkit.
        To overcome this limitation, we partition long audio files.
        
        %% This figure illustrates audio partitioning.
        %% Uncomment if needed (or, if you have text spacing to spare)
        %\begin{figure}[t]
        %\centering
        %\input{fig/partitioning}
        %\vspace{-2mm}
        %\caption{Partitioning of longer audio files to smaller parts.}
        %\vspace{-4mm}
        %\label{fig:partitioning}
        %\end{figure}
        
        %% More detailed description of partitioning, uncomment if needed.
        %Partitioning is done by splitting the audio into smaller blocks, performing inference on the parts, and then concatenating the posteriors to perform CTC segmentation.
        %A long audio file of length $l_{\text{full}}$ samples is partitioned into parts with length $l_{\text{cut}}$.
        %The maximum length of the last part is chosen to $l_{\text{last}}\leq  1.25\cdot l_{\text{cut}}$ to prevent too short audio parts at the end.
        %For our experiments, $l_{\text{last}}$ includes up to $320$s of audio.
        %As abrupt cuts in the audio cause distortions during inference, we add an overlap of length $l_{\text{overlap}}$ to each side of the block.
        %By inspection of CTC posteriors, an overlap of at least $600$ms is required to reduce distortions and impact on scoring to an acceptable level.
        %we chose a value corresponding to $1$s of audio. % comment: this value results in different durations depending on model, see comment about confidence score length above
        %Before concatenating the inferred posteriors of the parts, the CTC layer outputs that relate to the overlaps are omitted.
        %All lengths were chosen as multiples of the samples--to--posteriors ratio to preserve timing information and ensure the correct shape of the concatenated CTC posterior tensor.

        Partitioning is done by splitting the audio into smaller blocks, performing inference on the parts, and then concatenating CTC posteriors.
        Maximum block sizes were chosen depending on the memory consumption of the model and available memory;
        the last block may be $25\%$ longer to avoid too short blocks.
        As abrupt cuts in the audio cause distortions during inference, we add an overlap to each side of the block; overlapping posteriors are later omitted for scoring.
        By inspection of CTC posteriors, an overlap of at least $600$ms is required to reduce impact of partitioning on scoring to an acceptable level.
        All lengths were chosen as multiples of the samples--to--posteriors ratio to preserve timing information and ensure the correct shape of the concatenated CTC posterior tensor.

    \subsection{Specific cleansing for speaker verification} \label{sec:asv} \vspace{-1mm}
        Unlike an ASR corpus, ASV requires high-quality speaker labels. Therefore, we propose an unsupervised method for only extracting monologue videos (i.e., single-speaker videos) from the obtained videos. Furthermore, we remove videos with text-to-speech (TTS) voices, which have different characteristics from natural speech.
            
        \drawfig{t}{0.60\linewidth}{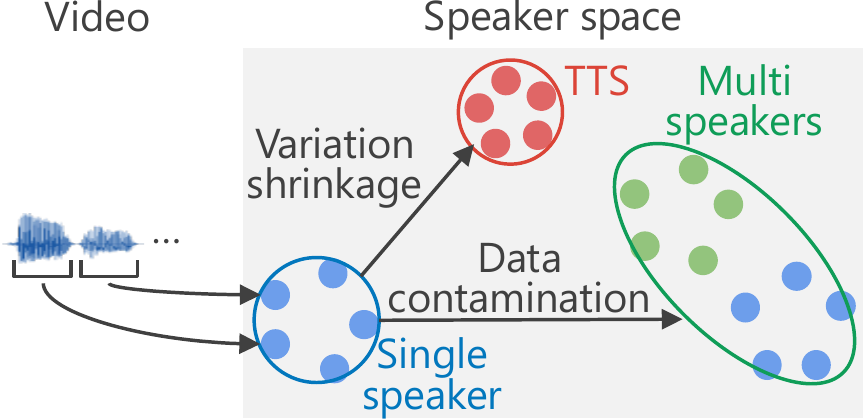}
        {Comparison of distributions in speaker space.}

        %\vspace{-1mm}
        %\subsubsection{Removing non-speech and too-short videos}  \vspace{-1mm}
        \textbf{Removing non-speech and too-short videos.}
        First, we delete videos without speech. Unlike the corpus for ASR, there is no need to align subtitle and speech. Therefore, we simply use voice activity detection (VAD) here. We applied VAD to sections trimmed out based on the subtitles and used only sections mainly consisting of speech. Also, too-short videos are deleted. This is for robustly calculating the intra-video statistics described below.
        
        %\vspace{-1mm}
        %\subsubsection{Evaluating intra-video variation in speaker space}  \vspace{-1mm}
        \textbf{Evaluating intra-video variation in speaker space.}
        To extract single-speaker videos, we compute the speaker variability in the video. \Fig{fig/single-speaker-collection} illustrates the concept. We use the $d$-vector~\cite{variani14dvecs}, deep learning-based speaker representation, which is extracted using pre-trained models. The $d$-vector is first calculated for each utterance. Then, the variance of the $d$-vectors is calculated within the video. We expect that the variance of TTS voices becomes smaller than the single-speaker voices because the TTS voice has no fluctuation among utterances. Also, if the different speakers' voices are contaminated (i.e., multi-speaker video), the variance will become larger than the single-speaker video. Therefore, we can eliminate TTS videos and multi-speaker videos by setting the appropriate threshold to the variance. For implementation, the $d$-vector is reduced to a lower dimension by t-SNE~\cite{maaten08tsne}, and the determinant of its covariance matrix is obtained.
            
            %%% ASRU %%%
            %To extract single-speaker videos, we compute the speaker variability in the video. \Fig{fig/single-speaker-collection} illustrates the concept of evaluating intra-video variation in speaker space. We use the $d$-vector~\cite{variani14dvecs}, deep learning-based speaker representation, which is extracted using pre-trained models. The $d$-vector is first calculated for each utterance trimmed out based on the subtitles. Then, the variance of the $d$-vectors is calculated within the video. We expect that the variance of TTS voices becomes smaller than the single-speaker voices because the TTS voice has no fluctuation between utterances. Also, if the different speakers' voices are contaminated (i.e., multi-speaker video), the variance will become larger than the single-speaker video. Therefore, we can eliminate TTS videos and multi-speaker videos by setting the appropriate threshold to the variance. For implementation, the $d$-vector is reduced to a lower dimension by t-SNE~\cite{maaten08tsne}, and the determinant of its covariance matrix is obtained.
        
        %\vspace{-1mm}
        %\subsubsection{Grouping videos by YouTube Channel ID} \vspace{-1mm}
        \textbf{Grouping videos by YouTube Channel ID.}
        It is necessary to prevent the same speaker from being counted as different speakers. We obtain and use the YouTube Channel ID of each video. Single-speaker videos with the same Channel ID are grouped together and considered as a unique speaker.

% Section 3: Experimental evaluation
\vspace{-2mm}
\section{Experimental evaluation} \label{sec:experiment}
\vspace{-2mm}

    \subsection{Evaluation in data collection} \vspace{-1mm}
        The period of data collection was between February and April of 2021. From the data collection results listed in \Table{corpus_spec}, we can describe 1) $5.09$ videos are found for each search term and 2) $0.92$~\% of videos have manual subtitles, and $41.7$~\% have automatic subtitles. In the end, we obtained approximately $10,000$ hours of speech data from $110,000$ YouTube videos. % Most of the videos are less than 5 minutes long, and the average duration is 3.8 minutes. Also, almost all utterances are shorter than 25 seconds because the subtitles are synchronized with the video. 
        
        %%% ASRU %%%
        % The video data was collected between February and April of 2021. Also, the Wikipedia timestamp used to create search terms was February 2021. \Table{corpus_spec} lists the number of search terms and videos we collected. From these numbers, we can describe that
        % \begin{itemize} \leftskip -5mm \itemsep -1mm
        %     \item $5.09$ videos are found for each search term.
        %     \item $0.92$~\% of videos have manual subtitles, and $41.7$~\% have automatic subtitles.
        % \end{itemize}
        % In the end, we obtained approximately 10,000 hours of speech data from 110,000 YouTube videos. \Fig{fig/duration} shows duration of videos and utterances. Most of the videos are less than 5 minutes long, and the average duration is 3.8 minutes. Also, a few videos are several hours long but these are very rare. On the other hand, almost all utterances are shorter than 25 seconds because the subtitles are synchronized with the video, but a few utterances are several hours long. 
 
        \begin{table}[t]
    \centering
    \caption{Results of data collection. Videos with automatic subtitles are not used in this paper, but the video ID is also opensourced.}
    \begin{tabular}{l|l}
        Retrieved entity                                & Value  \\ \hline
        \#search-terms                  & 2.34M terms \\
        \#videos found in the search    & 11.9M videos \\
        \#videos with manual subtitles  & 0.11M videos \\
        (\#videos with auto subtitles)  & 4.96M videos \\ 
    \end{tabular}
    \label{tab:corpus_spec}
    \vspace{-4mm}
\end{table}

% 多言語
% Librispeech
% Common Voice
% SPGISpeech
% Automatic Construction of a Large-Scale Speech Recognition Database Using Multi-Genre Broadcast Data with Inaccurate Subtitle Timestamps
% https://ieeexplore.ieee.org/stamp/stamp.jsp?tp=&arnumber=9414423
% https://arxiv.org/pdf/2104.04896.pdf
% https://arxiv.org/pdf/2006.08274.pdf と [10]

        \begin{table}[t]
    \centering
    \footnotesize
    \caption{Comparison of speech corpora of Japanese (upper half) and other rich-resource languages (lower half).}
    \begin{tabular}{c|cccc}
    Lang.            & Task          & Corpus                             & Open-source & Duration \\ \hline
    Ja            & ASR/ASV       & JNAS~\cite{itou99jnas}                & No        & 90       \\
    Ja            & ASR           & CSJ~\cite{maekawa00csj}               & No        & 600 \\
    Ja            & ASR           & LaboroTVspeech~\cite{ando21laborotvspeech}  & Yes       & 2,000 \\
    Ja            & ASR           & Common Voice~\cite{ardila20commonvoice} & Yes  & 2 \\ 
    Ja            & ASV           & Liveness~\cite{shiota15liveness}      & No        & 4 \\
    Ja            & ASR/ASV       & \textbf{JTubeSpeech (ours)}           & Yes       & 1,300/900 \\ \hline
    En             & ASR           & Librispeech~\cite{panayotov2015librispeech} & Yes  & 982 \\    
    En             & ASR           & Common Voice~\cite{ardila20commonvoice} & Yes  & 1,100 \\
    En             & ASR           & SPGISpeech~\cite{oneill21spgispeech} &  Yes       & 5,000 \\ 
    En             & ASR           & GigaSpeech~\cite{chen2021gigaspeech}  & Yes       & 10,000 \\
    En    & ASV           & VoxCeleb~\cite{nagrani19voxceleb}     & Yes       & 2,800 \\ 
    Zh             & ASR           & Common Voice~\cite{ardila20commonvoice} & Yes & 12 \\
%    Zh             & ASR           & HKUST~\cite{liu06hkust}               & No        & 200 \\
    Zh             & ASR           & AISHELL-2~\cite{du18aishell2}         & Yes       & 1,000 \\
    Zh             & ASV           & CN-Celeb~\cite{fan19cnceleb}          & Yes       & 1,000 \\
    \end{tabular}
    \label{tab:corpus_comparison}
    \vspace{-4mm}
\end{table}

        \Table{corpus_comparison} shows a comparison with the existing corpus. The duration of our corpus is a subset used in the ASR or ASV experiments described below. Among ASR corpora, our corpus is similar in size to LaboroTVspeech (Ja)~\cite{ando21laborotvspeech} and Common Voice (En)~\cite{ardila20commonvoice}. Also, among ASV corpora, it is the first open-source Japanese corpus and is similar in size to CN-Celeb (Zh)~\cite{fan19cnceleb}. 
        
        %%% ASRU %%%
        % \Table{corpus_comparison} shows a comparison with the existing corpus. The duration of our corpus is not all of the collected data described above but subsets used in the ASR and ASV experiments described below. Among ASR corpora, our corpus is similar in size to LaboroTVspeech (Japanese)~\cite{ando21laborotvspeech} and Common Voice (English)~\cite{ardila20commonvoice}. Also, among ASV corpora, it is the first open-source Japanese corpus and is similar in size to CN-Celeb (Chinese)~\cite{fan19cnceleb}. We are planning to further analyze the collected data and expand the corpus size.

    \subsection{Evaluation in speech recognition} \vspace{-1mm}
        %\subsubsection{Data cleaning} \vspace{-1mm}
        %\label{sec:data_cleaning}
        \textbf{Data cleaning.}
        The most important data pre-processing is to prune utterances that have the wrong transcriptions and fix the incorrect timing.
        We consistently employed CTC segmentation and CTC scoring, as described in Section \ref{sec:asr}.
        We used a pre-trained CTC model based on the ESPnet LaboroTVspeech~\cite{ando21laborotvspeech} recipe~\cite{Laborotvrecipe}. %\footnote{\url{https://github.com/espnet/espnet/tree/master/egs2/laborotv/asr1}}. 
        Table \ref{tab:asr_data_stats} summarizes the statistic of the various training and test sets.

        %As a pilot study, we used two subsets of our corpus:
        %\begin{itemize}
        %    \item "single\_speaker" (or "ss" in short): based on the single speaker subset, as described in Section \ref{sec:asv}
        %    \item "top\_15k" (or "15k" in short): extracted based on top 15,000 videos in terms of the average score for each video.
        %\end{itemize}
        As a pilot study, we used two subsets of our corpus:
        (1) "single\_speaker" (or "ss" in short) based on the single speaker subset, as described in Section \ref{sec:asv}, and
        (2) "top\_15k" (or "15k" in short) extracted based on top 15,000 videos in terms of the average score for each video.
        %The total file size of the single\_speaker and top\_15k subsets including the silence region amounts to 100GB and 300GB, respectively.
        Note that top\_15k is pruned only based on utterance confidence scores and it may contain multi speaker videos, unlike single\_speaker.
        Also a part of these subsets is overlapped.
        
        \drawfig{t}{0.85\linewidth}{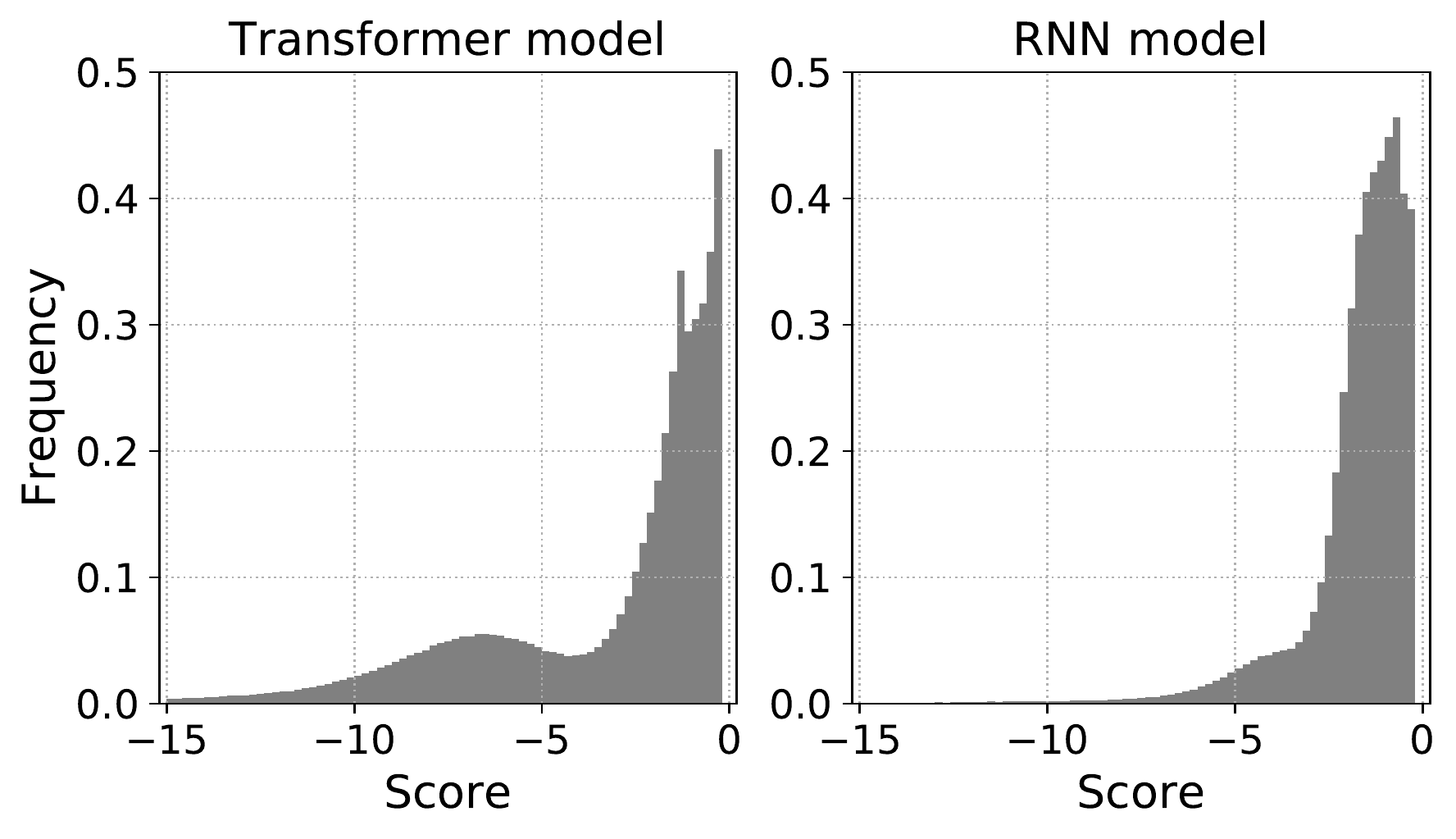}
        {Histograms of the score for different ASR models over all utterances of top\_15k.} %, with a scoring length that corresponds to $0.96$s of audio.}

        These subsets are further decomposed into the training and test sets, which will be explained in the next section.
        We also prepare the various training sets by changing the threshold value $\theta$ of the confidence score obtained by CTC segmentation for the single\_speaker and top\_15k subsets.
        To determine the threshold value, we investigate the distribution of the confidence score over all utterances in the top\_15k subset, as shown in \Fig{fig/hist_score_L20}.
        This distribution clearly shows that the peak of the distribution starts around -3.0 in both RNN and transformer based CTC models.
        Based on the observation, we regard the utterances located in the wide-based region as outlier data points, and use -3.0 as the lowest threshold for pruning in Table~\ref{tab:asr_data_stats}.
        The largest training set is obtained by combining the single\_speaker and top\_15k subsets (train\_ss\_15k).
        % It contains more than 17 thousand videos, 1.2 million utterances, and 1.3 thousand hours\footnote{We can further scale up our training data up to approximately five times more if we use the entire corpus.}.
        
        %\subsubsection{Test set design} \vspace{-1mm}
        %\label{sec:test_set_design}
        \textbf{Test set design.}
        %We carefully designed a test set for ASR evaluation from our corpus.
        Focusing on the "single\_speaker" videos reduces the cost of manual process of verifying audio and transcriptions.
        The rest of the procedure is as follows:
            (1) Select 1,621 videos from the "single\_speaker" videos that include "easy" utterances scored more than -0.3 threshold value based on CTC.
            (2) Randomly pick up 324 videos, around 20\%, and use them as a test video set, which has 3,396 easy utterances in total.
            (3) Manually listen utterances and identify 1,614 utterances having correct transcriptions.
            (4) Split them into the development \textbf{dev\_easy\_jun21} and evaluation \textbf{eval\_easy\_jun21} sets, which have 785 and 829 utterances, respectively.
        %First, we only focused on the "single\_speaker" videos to reduce the cost of manual process of verifying the audio and corresponding correct transcriptions.
        %The rest of the procedure is as follows:
        %\begin{enumerate}
        %    \item Select 1,621 videos from the "single\_speaker" videos that include "easy" utterances scored more than -0.3 threshold value based on CTC.
        %    \item Randomly pick up 324 videos (around 20\%) and use them as a test video set, which has 3,396 easy utterances in total.
        %    \item Manually listen utterances and identify 1,614 utterances having correct transcriptions.
        %    \item Split them into the development (\textbf{dev\_easy\_jun21}) and evaluation (\textbf{eval\_easy\_jun21}) sets, which have 785 and 829 utterances, respectively.
        %\end{enumerate}
        We also made an additional test set with "normal" utterances\footnote{
        The definition of "easy" and "normal" test sets are determined based on the CTC score objectively. Our listening process and later ASR experiments confirmed that this categorization is reasonable.
        } scored more than $-1.0$ threshold value.
        We fixed to use the same test video set as we defined before, and performed steps 3 and 4.
        We made \textbf{dev\_normal\_jun21} and \textbf{eval\_normal\_jun21} sets, which have 1,036 and 834 utterances, respectively.
        %We will perform the above procedure for the different threshold value or non single\_speaker subsets to create more difficult (e.g., hard) test sets in the future.

        \begin{table}[t]
            \centering
            \footnotesize
            \caption{Training and test data statistics for the ASR task. $\theta$ is a threshold used to prune bad utterances based on the CTC score.}
            \label{tab:asr_data_stats}
            \begin{tabular}{c|rrrr}
            & $\theta$ & \# videos & \# utts & hours \\
            \hline
            dev\_easy\_jun21       & -0.3 & 110   & 785     & 0.7 \\
            eval\_easy\_jun21      & -0.3 & 106   & 829     & 0.7 \\
            dev\_normal\_jun21     & -1.0 & 128   & 1,036   & 1.1 \\
            eval\_normal\_jun21    & -1.0 & 129   & 834     & 0.8 \\
            \hline
            train\_single\_speaker & -0.3 & 1,297 & 14,797  & 12.7 \\
            train\_single\_speaker & -0.5 & 1,792 & 26,209  & 24.2 \\
            train\_single\_speaker & -1.0 & 2,906 & 66,563  & 71.9 \\
%            train\_single\_speaker & -2.0 & 4,068 & 186,733 & 227.4 \\
            train\_single\_speaker & -3.0 & 4,342 & 285,846 & 362.0 \\
            \hline
            train\_top\_15k & -3.0 & 14,418 & 1,048,699 & 1087.1 \\
            \hline
            train\_ss\_15k & -3.0 & \textbf{17,761} & \textbf{1,270,124} & \textbf{1376.9} \\
            \end{tabular}
            \vspace{-4mm}
        \end{table}

        %\subsubsection{ASR models} \vspace{-1mm}
        \textbf{Experimental results.}
        We used an ESPnet state-of-the-art conformer model \cite{gulati20conformer,guo2021recent} based on hybrid CTC/attention architectures \cite{watanabe20202020}.
        %The conformer encoder has 12 conformer blocks with a kernel size of 31 and the transformer decoder has 8 blocks.
        %Encoder and decoder blocks each have 512 attention dimensions, 8 attention heads, and $2,048$ feed forward dimensions.
        %We used SpecAugment during training \cite{park2019specaugment}.
        %Inference was performed without LM.
        %We did not use LM shallow fusion during inference.
        The detailed configuration can be found in the ESPnet JTubeSpeech recipe. %\footnote{\url{https://github.com/espnet/espnet/pull/3311}}. % OR: ~\cite{jtubespeechpullrequest}

        %\subsubsection{Experimental results} \vspace{-1mm}
        %\textbf{Experimental results.}
        \Fig{fig/cer_hour2} shows the ASR performance for various amounts of training data by changing the score threshold $\theta$.
        We confirmed that the normal test set was more difficult than the easy test set.
        %(the error rate was increased by 1.6 to 2.1 times more).
        This result validates our design of the test set by controlling the difficulties based on the score.
        As regards the amounts of training data, if we increase the amount of training data by reducing $\theta$, the training data may contain more noisy transcriptions or distorted audio segments.
        Nevertheless, the ASR performance was improved, and the final CERs were $5.2\%$ in eval\_easy\_jun21 and $10.7\%$ in eval\_normal\_jun21.
        These CER ranges are similar to other Japanese ASR benchmarks, e.g., $4-6\%$ in CSJ \cite{maekawa00csj} and $13\%$ in LaboroTVspeech \cite{ando21laborotvspeech}.
        %Thus, we can conclude that our ASR benchmark is reasonable as our initial step of the corpus design.
        We also list the largest training data case (ss\_15k\_1376h) by combining the single\_speaker subset and top\_15k subsets.
        %This data was obtained by combining the single\_speaker subset and the top\_15k subset, which contains multi-speaker recordings. %, as described in Section \ref{sec:data_cleaning}. 
        The performance was improved in most cases except for dev\_easy\_jun21.
        %possibly due to the mismatch between single speaker and multi-speaker recordings.
        %The eval\_normal\_jun21 CER was improved from $10.7\%$ to $10.0\%$.
        This result confirms that 
        %while we need more careful cleaning to maintain the clean speech recognition task performance, 
        more training data is generally helpful to improve the performance of more challenging data.
        
        \drawfig{t}{0.98\linewidth}{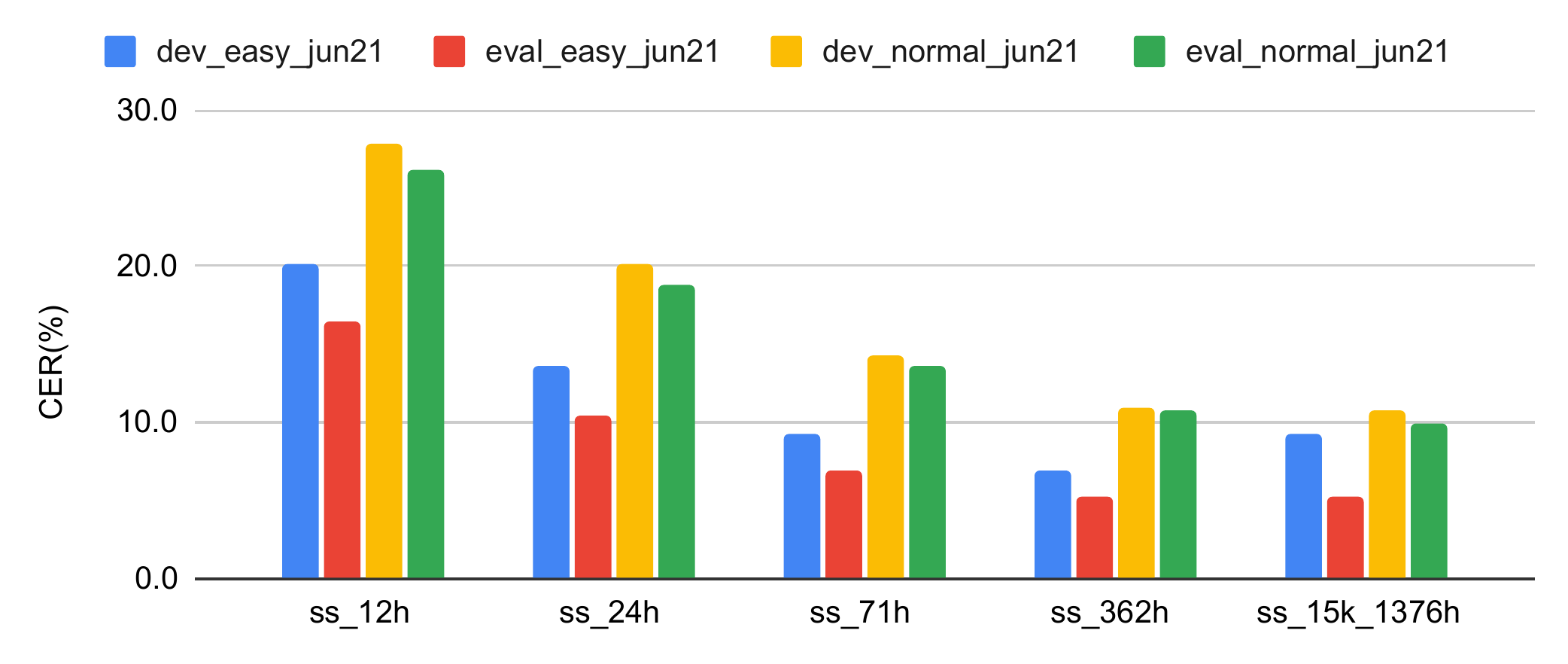}
        {ASR performance for various amounts of training data (12, 24, 71, 362, and 1376 hours) by changing the score threshold and combining multi-speaker recordings. 
        %See Table \ref{tab:asr_data_stats} for the detailed training data information.
        } 
    
        Finally, we evaluate the effectiveness of CTC segmentation, as discussed in Section \ref{sec:asr}.
        We used train\_single\_speaker with $-1.0$ score threshold (71.9 hours) and prepared the corresponding training data based on the original YouTube timings only with CTC scoring.
        Table \ref{tab:ctc_seg} shows that the CTC segmentation significantly improved the performance and shows the effectiveness of the re-aligning for the YouTube audio data, as suggested by \cite{chen2021gigaspeech}.
    
        \begin{table}[t]
            \centering
            \footnotesize
            \caption{The effectiveness of CTC segmentation.}
            \label{tab:ctc_seg}
        \begin{tabular}{l|cr|cr}
                                         & dev\_easy                     & \multicolumn{1}{c|}{dev\_normal} & dev\_easy                    & \multicolumn{1}{c}{dev\_normal} \\ \hline
        \multicolumn{1}{r|}{original timing} & \multicolumn{1}{r}{11.5} & 16.5                        & \multicolumn{1}{r}{9.2} & 15.7                       \\
        \multicolumn{1}{r|}{CTC segmentation}  & \multicolumn{1}{r}{\textbf{9,2}}  & \textbf{14.3}                       & \multicolumn{1}{r}{\textbf{6.9}} & \textbf{13.6}                      
        \end{tabular}
        \vspace{-4mm}
        \end{table}    
    
    \vspace{-1mm}
    \subsection{Evaluation in speaker verification} \vspace{-1mm}
        \subsubsection{Data cleansing} \vspace{-1mm}
        We used py-webrtcvad\footnote{\url{https://github.com/wiseman/py-webrtcvad}} for VAD and a pre-trained model\footnote{\url{https://github.com/yistLin/dvector}} for extracting $d$-vectors. The variation in speaker space was computed from more than 10 utterances for each video. As a pilot study, we used randomly selected 35,000 videos (approx. 30\% of total). \Fig{fig/tsne_score_rev} shows two levels of rapid increase: near 0 and near 8 to 9 on the x-axis. Following the concept in \Fig{fig/single-speaker-collection}, we set a threshold around these levels and assign ``TTS,'' ``single speaker,'' or ``multi speakers,'' class to the videos.
        
        %%% ASRU %%%
        % We used py-webrtcvad\footnote{\url{https://github.com/wiseman/py-webrtcvad}} for VAD and an open-sourced pre-trained model\footnote{\url{https://github.com/yistLin/dvector}} for extracting $d$-vectors. The intra-video variation in speaker space was computed from more than 10 utterances for each video. As a pilot study, we used randomly selected 35,000 videos (approx. 30\% of total).

        %%% ASRU %%%
        % \Fig{fig/tsne_score_rev} shows the log determinant of a covariance matrix of $d$-vectors in each video (i.e., intra-video variation). In this figure, we can see two levels of rapid increase. The first level is near 0 on the x-axis, which means that videos with extremely low variance are included in the collected data. The second level is near 8 to 9 on the x-axis. Following the concept in \Fig{fig/single-speaker-collection}, we set a threshold around these levels and assign ``TTS,'' ``single speaker,'' or ``multi speakers,'' class to the videos, starting from the lowest value.
    
        \drawfig{t}{0.7\linewidth}{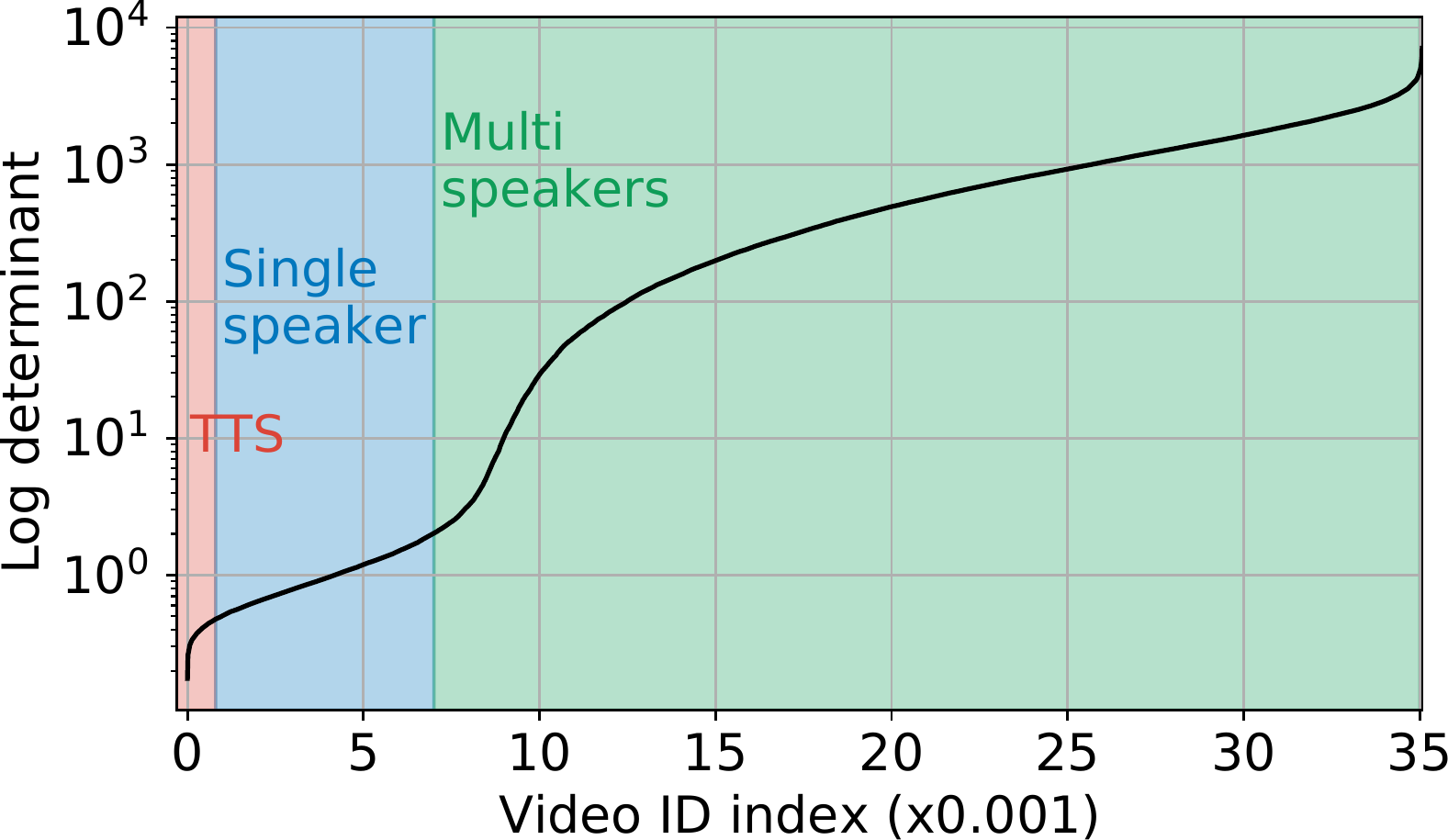}
        {The intra-video variation in speaker space}    

        We quantitatively evaluate the authenticity of this classification. We randomly extracted 100 videos from each class and annotated a true class label. 
        %Multi-TTS videos (i.e., multiple TTS systems) and overdubbed-voice videos (e.g., overdubbed singing voices) were annotated as ``multi speakers.'' 
        Multi-TTS videos and overdubbed-voice videos were annotated as ``multi speakers.'' 
        To accurately find TTS videos, a TTS specialist participated in this annotation. \Table{video_type} shows that each class has their suitable videos. Especially, the ``single speaker'' class does not contain multi-speaker videos. Therefore, our method works for choosing the single-speaker videos. Some single-speaker videos leak to the ``TTS'' class, but this impact is limited because the ``TTS'' class has only a few videos.
        
        %%% ASRU %%% 
        % \Table{video_type} lists the result. It is observed that each class has their suitable videos. Especially, the ``single speaker'' class does not have multi-speaker videos. Therefore, we can demonstrate that our method works for eliminating the multi-speaker videos and choosing the single-speaker videos. Some single-speaker videos leak to the ``TTS'' class, but this impact is limited because the ``TTS'' class has only a few videos.

        \begin{table}[t]
    \centering
    \footnotesize
    \caption{Comparison of classified and annotated video types.}
    \begin{tabular}{c|c|c|c}
    Classified \textbackslash Annotated   & TTS    & single speaker  & multi speakers \\ \hline
    TTS                     & 20     & 80       & 0 \\
    single speaker          & \textbf{5}      & \textbf{95}       & \textbf{0} \\
    multi speakers          & 1      & 36       & 63 \\
    \end{tabular}
    \vspace{-4mm}
    \label{tab:video_type}
\end{table}

        \subsubsection{ASV dataset design and models} \vspace{-1mm}
        %For considering the usage of the same evaluation sets as ASR, we designed the evaluation trials from the ASR evaluation set (\textbf{eval\_easy\_jun21)}.  
        From the single-speaker videos, 92 unique speakers were selected as the enrollment and testing speakers. For the training and development datasets, 1,795 unique speakers were selected.
        %Then, we selected 1,795 unique speakers for a training dataset and development one %from the single-speaker videos except for the videos of the 92 speakers. Consequently, 3498 videos were used for training and development datasets. 
        In the training, 127,997 and 25,392 utterances were used for training and development datasets, respectively. 
        In the testing, we designed an enrollment utterance and a testing one were picked up from the same video but different segments. %For picking up the test utterances accurately, the manual subtitle information was used. 
        In the testing, we performed 25,392 trial pairs, included 276 correct pairs and 25,116 incorrect pairs.

        %\subsubsection{ASV models} \vspace{-1mm}      
        We used a speaker embedding network, which was built with four convolutional neural networks, pooling layers and two full connected layers. %The output of the final layer was regarded as a speaker embedding vector. 
        As the input feature, 40-order Mel-Frequency Cepstrum Coefficients with 25-ms frame length and half overlapped shift were used. The speaker-embedding vector was 512 dimensions. The evaluation metric is equal error rate (EER) was used for our evaluation by comparing the cosine distance.
        %The speaker embedded vectors extracted from the enrollment utterances and the testing utterance were compared by the cosine distance similarity. Equal error rate (EER) was used for our evaluation.

        \subsubsection{ASV performance} \vspace{-1mm}
        We performed ASV evaluations, and the EER of our ASV system with JTubeSpeech was \textbf{10.9\%}.  Our system was regarded as the speaker embedding system based on the CNN-based model. Even though such a simple system, this result had the similar performance to the CNN-based model (VGG-M) trained with Voxceleb1~\cite{nagrani19voxceleb}.
        As a pilot study, we designed a simple benchmark. For example, the result was obtained with the small model and probabilistic linear discriminant analysis (PLDA) scoring and data augmentation techniques were not performed. %Since the trial pairs of the correct cases were picked up from the same video in the correct pair cases, some trial part seemed to be easy. 
        However, from this simple benchmark, we could show the techniques of data cleaning for selecting a unique speakers and 
        constructing ASV systems performed well. Consequently, we established the Japanese ASV systems with large-scale open source media in the first time.
%        because of the dataset came from Youtube videos, there were hevy BGM and sound effects. The 

%        Table \ref{tab:asv_results} shows the EER of our ASV system with JTubeSpeech. As a pilot study, we designed a simple benchmark. For example, the result was obtained with a small model and probabilistic linear discriminant analysis (PLDA) scoring and data augmentation techniques were not performed. Since the trial pairs were picked up from the same video in the correct pair cases, some part seemed to be easy.         

%        To fully utilize YouTube videos for estimating reliable ASV systems, data augmentation and speech enhancement techniques and other data cleaning schemes will be required. In addition, as shown in Table \ref{tab:video_type}, some TTS speakers remained in the single speakers dataset. We will discuss the influences of including TTS speakers and how to eliminate such speakers. To keep the reproducibility of our ASV system, we will add the ASV recipe based on the ESPnet style in the future.

\begin{comment}
    \begin{table}[t]
        \centering
        \caption{EER (\%) of ASV evaluation.}
        \label{tab:asv_results}
        \begin{tabular}{c|r}
        & eval\_easy\_jun21  \\
        \hline
        Spk embeddings / cosine dist. & 10.9\% 
        \end{tabular}
    \end{table}  
\end{comment}

% Section 4: Conclusion
\vspace{-2mm}
\section{Conclusion} \label{sec:conclusion}
\vspace{-2mm}
In this paper, we propose a speech corpus construction strategy and build $1,300$ and $900$ hours data for Japanese ASR and ASV. One of our future directions is to extend the corpus to multiple languages. 

%In this paper, we propose a speech corpus construction strategy and build $1,300$ and $900$ hours data for Japanese ASR and ASV. First, we retrieved audio and subtitles from YouTube videos with manual subtitles. Then, we constructed corpora for ASR and ASV by CTC-based cleaning and speaker variation-based cleaning. One of our future directions is to extend the corpus to multiple languages. Our method (and open-sourced scripts) is almost language-independent, so it can be easily applied to languages other than Japanese. Another direction is the use of automatic subtitles, which are 45 times larger than manual subtitles.

% acknowledgement
%\section{Acknowledgement}

%%%% moved to page 1 %%%%
% \hspace{-5mm} {\footnotesize \textbf{Acknowledgement:}
% We would like to thank Hiromasa Fujihara and the GigaSpeech team, especially Guoguo Chen and Shuzhou Chai for their valuable comments.
% This work used the Human Language Technology Center of Excellence (HLTCOE) cluster at Johns Hopkins University and the Extreme Science and Engineering Discovery Environment (XSEDE) \cite{towns2014xsede}, which is supported by National Science Foundation grant number ACI-1548562. Specifically, it used the Bridges system \cite{nystrom2015bridges}, which is supported by NSF award number ACI-1445606, at the Pittsburgh Supercomputing Center (PSC). This work is also supported by JSPS KAKENHI 17H06101, 19H01116, 19K20271, 21H04900, ROISDS-JOINT (030RP2021) to S. Shiota and the SECOM Science and Technology Foundation.
% }

% bib
\newpage
%{\footnotesize
% \bibliographystyle{bib/IEEEbib}
% \bibliography{bib/refs}
\printbibliography

@misc{fan19cnceleb,
  title={{CN-CELEB}: a challenging {C}hinese speaker recognition dataset},
  author={Yue Fan and Jiawen Kang and Lantian Li and Kaicheng Li and Haolin Chen and Sitong Cheng and Pengyuan Zhang and Ziya Zhou and Yunqi Cai and Dong Wang},
  year={2019},
  eprint={1911.01799},
  archivePrefix={arXiv},
  primaryClass={eess.AS}
}

@inproceedings{liu06hkust,
  title={{HKUST/MTS}: A very large scale {M}andarin telephone speech corpus},
  author={Liu, Yi and Fung, Pascale and Yang, Yongsheng and Cieri, Christopher and Huang, Shudong and Graff, David},
  booktitle={International Symposium on Chinese Spoken Language Processing},
  pages={724--735},
  year={2006},
  organization={Springer}
}

@article{itou99jnas,
author = {K. Itou and M. Yamamoto and K.Takeda and T. Takezawa and T. Matsuoka and T. Kobayashi and K. Shikano and S. Itahashi},
journal = {Journal of the Acoustical Society of Japan (E)},
title = {{JNAS}: {J}apanese speech corpus for large vocabulary continuous speech recognition research},
volume = {20},
number = {3},
pages = {199--206},
year = {1999},
month = {May},
}

@Article{nagrani19voxceleb,
  author       = "Arsha Nagrani and Joon~Son Chung and Weidi Xie and Andrew Zisserman",
  title        = "Voxceleb: Large-scale speaker verification in the wild",
  journal      = "Computer Science and Language",
  year         = "2019",
  publisher    = "Elsevier",
}

@inproceedings{ardila20commonvoice,
    title = "{Common Voice}: A Massively-Multilingual Speech Corpus",
    author = "Ardila, Rosana  and
      Branson, Megan  and
      Davis, Kelly  and
      Kohler, Michael  and
      Meyer, Josh  and
      Henretty, Michael  and
      Morais, Reuben  and
      Saunders, Lindsay  and
      Tyers, Francis  and
      Weber, Gregor",
    booktitle = "Proceedings of the 12th Language Resources and Evaluation Conference",
    month = may,
    year = "2020",
    address = "Marseille, France",
    publisher = "European Language Resources Association",
    url = "https://www.aclweb.org/anthology/2020.lrec-1.520",
    pages = "4218--4222",
}

@misc{oneill21spgispeech,
      title={{SPGISpeech}: 5,000 hours of transcribed financial audio for fully formatted end-to-end speech recognition}, 
      author={Patrick K. O'Neill and Vitaly Lavrukhin and Somshubra Majumdar and Vahid Noroozi and Yuekai Zhang and Oleksii Kuchaiev and Jagadeesh Balam and Yuliya Dovzhenko and Keenan Freyberg and Michael D. Shulman and Boris Ginsburg and Shinji Watanabe and Georg Kucsko},
      year={2021},
      eprint={2104.02014},
      archivePrefix={arXiv},
      primaryClass={cs.CL}
}

@inproceedings{maekawa00csj,
  title={Spontaneous Speech Corpus of {J}apanese.},
  author={Maekawa, Kikuo and Koiso, Hanae and Furui, Sadaoki and Isahara, Hitoshi},
  booktitle={Proc. LREC},
  pages={947--952},
  year={2000}
}

@inproceedings{shiota15liveness,
  title={Voice liveness detection algorithms based on pop noise caused by human breath for automatic speaker verification},
  author={Shiota, Sayaka and Villavicencio, Fernando and Yamagishi, Junichi and Ono, Nobutaka and Echizen, Isao and Matsui, Tomoko},
  booktitle={Sixteenth annual conference of the international speech communication association},
  year={2015}
}

@INPROCEEDINGS{ando21laborotvspeech,
  author={Ando, Shintaro and Fujihara, Hiromasa},
  booktitle={ICASSP 2021 - 2021 IEEE International Conference on Acoustics, Speech and Signal Processing (ICASSP)}, 
  title={Construction of a Large-Scale {J}apanese {ASR} Corpus on {TV} Recordings}, 
  year={2021},
  volume={},
  number={},
  pages={6948-6952},
  doi={10.1109/ICASSP39728.2021.9413425}}

@inproceedings{variani14dvecs,
 address = {Florence, Italy},
 author = {E. Variani and X. Lei and E. McDermott and I. L. Moreno and J. Gonzalez-Dominguez},
 booktitle = {Proc. ICASSP},
 title = {Deep neural networks for small footprint text-dependent speaker verification},
 month = {May},
 pages = {4080–-4084},
 year = {2014},
}

@inproceedings{snyder18xvector,
 address              = {Alberta, Canada},
 author               = {D. Snyder and D. Garcia-Romero and G. Sell and D. Povey and S. Khudanpur},
 booktitle            = {Proc. ICASSP},
 month                = {Apr.},
 pages                = {5329--5333},
 title                = {X-Vectors: robust {DNN} embeddings for speaker recognition},
 year                 = {2018},
}

@article{george11hybridasr,
  author={Dahl, George E. and Yu, Dong and Deng, Li and Acero, Alex},
  journal={IEEE Transactions on Audio, Speech, and Language Processing}, 
  title={Context-Dependent Pre-Trained Deep Neural Networks for Large-Vocabulary Speech Recognition}, 
  year={2012},
  volume={20},
  number={1},
  pages={30-42},
  doi={10.1109/TASL.2011.2134090}}

@inproceedings{graves2014e2easr,
  title={Towards end-to-end speech recognition with recurrent neural networks},
  author={Graves, Alex and Jaitly, Navdeep},
  booktitle={International conference on machine learning},
  pages={1764--1772},
  year={2014},
  organization={PMLR}
}

@inproceedings{gulati20conformer,
  author={Anmol Gulati and James Qin and Chung-Cheng Chiu and Niki Parmar and Yu Zhang and Jiahui Yu and Wei Han and Shibo Wang and Zhengdong Zhang and Yonghui Wu and Ruoming Pang},
  title={{Conformer: Convolution-augmented Transformer for Speech Recognition}},
  year=2020,
  booktitle={Proc. Interspeech 2020},
  pages={5036--5040},
  doi={10.21437/Interspeech.2020-3015},
  url={http://dx.doi.org/10.21437/Interspeech.2020-3015}
}

@misc{abuelhaija16youtube8m,
      title={{YouTube-8M}: A Large-Scale Video Classification Benchmark}, 
      author={Abu-El-Haija, Sami and Kothari, Nisarg and Lee, Joonseok and Natsev, Paul and Toderici, George and Varadarajan, Balakrishnan and Vijayanarasimhan, Sudheendra},
      year={2016},
      eprint={1609.08675},
      archivePrefix={arXiv},
      primaryClass={cs.CV}
}

@article{towns2014xsede,
  title={Xsede: Accelerating scientific discovery computing in science \& engineering, 16 (5): 62--74, sep 2014},
  author={Towns, John and Cockerill, Timothy and Dahan, Maytal and Foster, Ian and Gaither, Kelly and Grimshaw, Andrew and Hazlewood, Victor and Lathrop, Scott and Lifka, Dave and Peterson, Gregory D and others},
  journal={URL https://doi. org/10.1109/mcse},
  volume={128},
  year={2014}
}

@inproceedings{nystrom2015bridges,
  title={Bridges: a uniquely flexible HPC resource for new communities and data analytics},
  author={Nystrom, Nicholas A and Levine, Michael J and Roskies, Ralph Z and Scott, J Ray},
  booktitle={Proceedings of the 2015 XSEDE Conference: Scientific Advancements Enabled by Enhanced Cyberinfrastructure},
  pages={1--8},
  year={2015}
}

@InProceedings{ctcsegmentation,
author="K{\"u}rzinger, Ludwig
and Winkelbauer, Dominik
and Li, Lujun
and Watzel, Tobias
and Rigoll, Gerhard",
editor="Karpov, Alexey
and Potapova, Rodmonga",
title="CTC-Segmentation of Large Corpora for German End-to-End Speech Recognition",
booktitle="Speech and Computer",
year="2020",
publisher="Springer International Publishing",
address="Cham",
pages="267--278",
isbn="978-3-030-60276-5"
}

@inproceedings{guo2021recent,
  title={Recent developments on espnet toolkit boosted by conformer},
  author={Guo, Pengcheng and Boyer, Florian and Chang, Xuankai and Hayashi, Tomoki and Higuchi, Yosuke and Inaguma, Hirofumi and Kamo, Naoyuki and Li, Chenda and Garcia-Romero, Daniel and Shi, Jiatong and others},
  booktitle={ICASSP 2021-2021 IEEE International Conference on Acoustics, Speech and Signal Processing (ICASSP)},
  pages={5874--5878},
  year={2021},
  organization={IEEE}
}

@article{watanabe20202020,
  title={The 2020 {ESPNet} update: New features, broadened applications, performance improvements, and future plans},
  author={Watanabe, Shinji and Boyer, Florian and Chang, Xuankai and Guo, Pengcheng and Hayashi, Tomoki and Higuchi, Yosuke and Hori, Takaaki and Huang, Wen-Chin and Inaguma, Hirofumi and Kamo, Naoyuki and others},
  journal={arXiv preprint arXiv:2012.13006},
  year={2020}
}

@Article{maaten08tsne,
  title   =  "Visualizing data using {t-SNE}",
  author  =  "L.v.d. Maaten and G. Hinton",
  journal =  "Journal of machine learning research",
  volume  =  "9",
  pages   =  "2579--2605",
  year    =  "2008"
}

@article{chen2021gigaspeech,
  title={{GigaSpeech}: An Evolving, Multi-domain ASR Corpus with 10,000 Hours of Transcribed Audio},
  author={Chen, Guoguo and Chai, Shuzhou and Wang, Guanbo and Du, Jiayu and Zhang, Wei-Qiang and Weng, Chao and Su, Dan and Povey, Daniel and Trmal, Jan and Zhang, Junbo and others},
  journal={arXiv preprint arXiv:2106.06909},
  year={2021}
}

@article{galvez2021peoplespeech,
  title={{The People’s Speech}: A Large-Scale Diverse {E}nglish Speech Recognition Dataset for Commercial Usage},
  author={Galvez, Daniel and Diamos, Greg and Torres, Juan Manuel Ciro and Achorn, Keith and Gopi, Anjali and Kanter, David and Lam, Max and Mazumder, Mark and Reddi, Vijay Janapa},
  year={2021},
  note={\url{https://openreview.net/forum?id=R8CwidgJ0yT}},
}

@INPROCEEDINGS{panayotov2015librispeech,
  author={Panayotov, Vassil and Chen, Guoguo and Povey, Daniel and Khudanpur, Sanjeev},
  booktitle={2015 IEEE International Conference on Acoustics, Speech and Signal Processing (ICASSP)}, 
  title={Librispeech: An ASR corpus based on public domain audio books}, 
  year={2015},
  volume={},
  number={},
  pages={5206-5210},
  doi={10.1109/ICASSP.2015.7178964}}

@ARTICLE{du18aishell2,
   author = {{Du}, J. and {Na}, X. and {Liu}, X. and {Bu}, H.},
   title = "{{AISHELL-2}: Transforming Mandarin ASR Research Into Industrial Scale}",
   journal = {ArXiv},
   eprint = {1808.10583},
   primaryClass = "cs.CL",
   year = 2018,
   month = Aug,
}

@misc{bakhturina2021hifi,
      title={Hi-Fi Multi-Speaker English TTS Dataset}, 
      author={Evelina Bakhturina and Vitaly Lavrukhin and Boris Ginsburg and Yang Zhang},
      year={2021},
      eprint={2104.01497},
      archivePrefix={arXiv},
      primaryClass={eess.AS}
}

@article{du2018aishell,
  title={Aishell-2: Transforming mandarin asr research into industrial scale},
  author={Du, Jiayu and Na, Xingyu and Liu, Xuechen and Bu, Hui},
  journal={arXiv preprint arXiv:1808.10583},
  year={2018}
}

@inproceedings{cieri2004fisher,
  title={The Fisher corpus: A resource for the next generations of speech-to-text.},
  author={Cieri, Christopher and Miller, David and Walker, Kevin},
  booktitle={LREC},
  volume={4},
  pages={69--71},
  year={2004}
}

@article{zhang2021nemo,
  title={NeMo Inverse Text Normalization: From Development To Production},
  author={Zhang, Yang and Bakhturina, Evelina and Gorman, Kyle and Ginsburg, Boris},
  journal={arXiv preprint arXiv:2104.05055},
  year={2021}
}

@inproceedings{trmal2017kaldi,
  title={The Kaldi OpenKWS System: Improving Low Resource Keyword Search.},
  author={Trmal, Jan and Wiesner, Matthew and Peddinti, Vijayaditya and Zhang, Xiaohui and Ghahremani, Pegah and Wang, Yiming and Manohar, Vimal and Xu, Hainan and Povey, Daniel and Khudanpur, Sanjeev},
  booktitle={Interspeech},
  pages={3597--3601},
  year={2017}
}

@misc{num2words,
    title={Num2Words: Modules to convert numbers to words.},
    note={\url{https://github.com/savoirfairelinux/num2words}},
    journal={GitHub},
    author={Dupras, Virgil and Grigaitis, Marius and Ogawa, Taro},
    year={2021},
    month={Jun}
}

@misc{ctcsegpython,
    title={Ctc-Segmentation: Segment an audio file and obtain utterance alignments. (python package)},
    note={\url{https://github.com/lumaku/ctc-segmentation}},
    journal={GitHub}, author={K\"urzinger, Ludwig and Winkelbauer, Dominik},
    year={2021},
    month={Jun}
}

@misc{Laborotvrecipe,
    title={Espnet/Egs2/Laborotv/Asr1 · espnet/espnet},
    note={\url{https://github.com/espnet/espnet/tree/master/egs2/laborotv/asr1}},
    journal={GitHub},
    author={Various authors},
    year={2021},
    month={Aug}
}
%}

\end{document}